%% file: kronfeld.tex
\documentclass[letterpaper]{jpconf}
\usepackage{bm}
\usepackage{graphicx}

\begin{document}
\title{Predictions with lattice QCD}

\author{Andreas S~Kronfeld}

\address{Fermi National Accelerator Laboratory,
	Batavia, Illinois 60510, USA}

\author{for the Fermilab Lattice, MILC, and HPQCD Collaborations}

\ead{ask@fnal.gov}

\begin{abstract}
\input a
\end{abstract}

\section{Introduction and Background}

\input 1

\section{Semileptonic $D$ Decays}

\input 2

\section{Leptonic $D$ Decays}

\input 3

\section{Mass of the $B_c$ Meson}

\input 4

\section{Conclusions}

\input 5

\ack

This work has been supported in part by the U.S. National Science
Foundation, the Office of Science of the U.S. Department of Energy
(DOE), and the U.K. Particle Physics and Astronomy Research Council.
Fermilab is operated by Universities Research Association Inc., under
contract with the DOE.
Software development and hardware protyping were supported by SciDAC.

\section*{References}

\input r
\end{document}

%% file: a.tex
In recent years, we used lattice QCD to calculate some quantities that
were unknown or poorly known.
They are the $q^2$ dependence of the form factor in semileptonic $D\to
Kl\nu$ decay, the leptonic decay constants of the $D^+$ and $D_s$
mesons, and the mass of the $B_c$ meson.
In this paper, we summarize these calculations, with emphasis on their
(subsequent) confirmation by measurements in $e^+e^-$, $\gamma p$ and
$\bar{p}p$ collisions

%% file: 1.tex
The central theme of elementary particle physics is to find new
interactions of matter, energy, space and time.
When the matter in question is the quarks, one is faced with quark
confinement: quarks never appear freely; they are always bound inside
hadrons---baryons like the proton, or mesons like the pion or kaon.
In the Standard Model of elementary particles, confinement is a
phenomenon of quantum chromodynamics (QCD), the gauge theory of the
strong force.

Since only hadrons can be detected, the effects of quark confinement
must be calculated with QCD, before the experimental data
can be interpreted in terms of quarks.
In many cases, the best technique for doing the calculations is to
formulate QCD on a space-time lattice.
Lattice QCD and the Feynman path integral reduce the problem of to an
integral whose dimension scales as~$N^4$, where $N$ is the (linear)
lattice size.
The problem cries out for supercomputing.

In recent years, lattice QCD has reached the stage where many
calculations of hadron masses, mass splittings and operator matrix
elements agree with experimental measurements.
The key has been the inclusion of sea quarks, which are pairs of virtual
quarks swirling around inside hadrons.
The progress has been especially striking~\cite{Davies:2003ik} when
the sea quarks are implemented as staggered quarks,
using an action designed to reduce discretization effects.

One ingredient of these calculations is controversial.
Sea quarks are always computationally demanding, although staggered
quarks are by far the fastest.
Staggered quarks introduce some extra unwanted quarks, however.
The computer algorithms~\cite{Gottlieb:1987mq} and subsequent analysis
of the numerical data~\cite{Aubin:2003mg} remove them, but do so
differently for valence and sea quarks.
The difference can lead to violations of unitarity.
In the cases discussed here, it is plausible that such effects are small, 
but a proof is not yet at hand~\cite{Durr:2005ax}.

Less controversial is the treatment of heavy quarks.
In practice, the lattice spacing is not small enough to resolve the 
Compton wavelength of charmed and $b$ quarks.
Fortunately, chromodynamics at this length scale is simple enough to
factor it out from the computer simulation, and several methods
exist~\cite{Kronfeld:2003sd}.
Nevertheless, it is good to have check.

In this paper, we discuss three topics:
the normalization and $q^2$-dependence of the
$D\to Kl\nu$ form factor;
the decay constants of the $D^+$ and $D_s$ mesons;
and the mass of the $B_c$ meson.
Each of these lattice-QCD calculations was subsequently confirmed by
experimental measurements, satisfying a long-standing demand of
experimental physicists~\cite{Shipsey:2004wz}.
The quantities discussed here were ideal candidates: they are
straightforward to compute; they test the controversial aspects in
complementary ways; and the first ``good'' experimental measurements
were expected on the same time scale.
The success of the predictions is extremely encouraging.
In particular, the calculations for $D$ mesons are, in lattice QCD,
similar to those for $B$ mesons, whose $b$ quarks are considered likely
to exhibit new, non-Standard interactions.

%% file: 2.tex
Semileptonic decays such as $D\to Kl\nu$ proceed as follows.
A~quark (in this case, a charmed quark) emits a virtual $W$ boson,
thereby turning into a quark of a different flavor (in this case, a
strange quark).
The $W$ immediately disintegrates into a lepton-neutrino ($l\nu$) pair.
The rate depends on $q^2$, which is the invariant-mass-squared of~$l\nu$.
Some of the $q^2$ dependence stems from QCD through a function 
called a form factor (in this case, denoted $f_+(q^2)$).
The momentum transfer $q^2$ falls in the range 
$0\leq q^2\leq q^2_{\rm max}=(m_D-m_K)^2$.
In lattice QCD, discretization effects are smallest when the spatial
momentum~$\bm{p}$ of the kaon is small, which puts $q^2$ close
to~$q^2_{\rm max}$.

Experiments usually measure the branching fraction and quote
the normalization $f_+(0)$, after making assumptions about the
$q^2$~dependence.
While our results were still preliminary~\cite{Okamoto:2003ur},
experimental results came out for the normalization of
$D\to Kl\nu$~\cite{Ablikim:2004ej}
and $D\to\pi l\nu$~\cite{Huang:2004fr}.
The agreement with our final results~\cite{Aubin:2004ej} is excellent.
For example, we find $f_+^{D\to K}(0)=0.73(3)(7)$~\cite{Aubin:2004ej}
while the BES Collaboration measures 
$f_+^{D\to K}(0)=0.78(5)$~\cite{Ablikim:2004ej}.

In principle, the shape of the form factors can be computed
directly in lattice QCD.
In practice, we calculated at a few values of $\bm{p}$ and used a fit to
the Ansatz of Be\'cirevi\'c-Kaidalov (BK)~\cite{Becirevic:1999kt} to fix
the $q^2$ dependence.
It was important, therefore, to measure the $q^2$ dependence
experimentally.
In photoproduction of charm off fixed nuclear targets, the FOCUS
Collaboration was able to collect high enough statistics to trace out
the $q^2$ distribution of the decay~\cite{Link:2004dh}.
This setup does not yield an absolutely normalized branching ratio, so
one is left to compare $f_+(q^2)/f_+(0)$.

\begin{figure}[b]
	\centering
	(a)\hspace*{-0.5em}\includegraphics[width=0.45\textwidth]{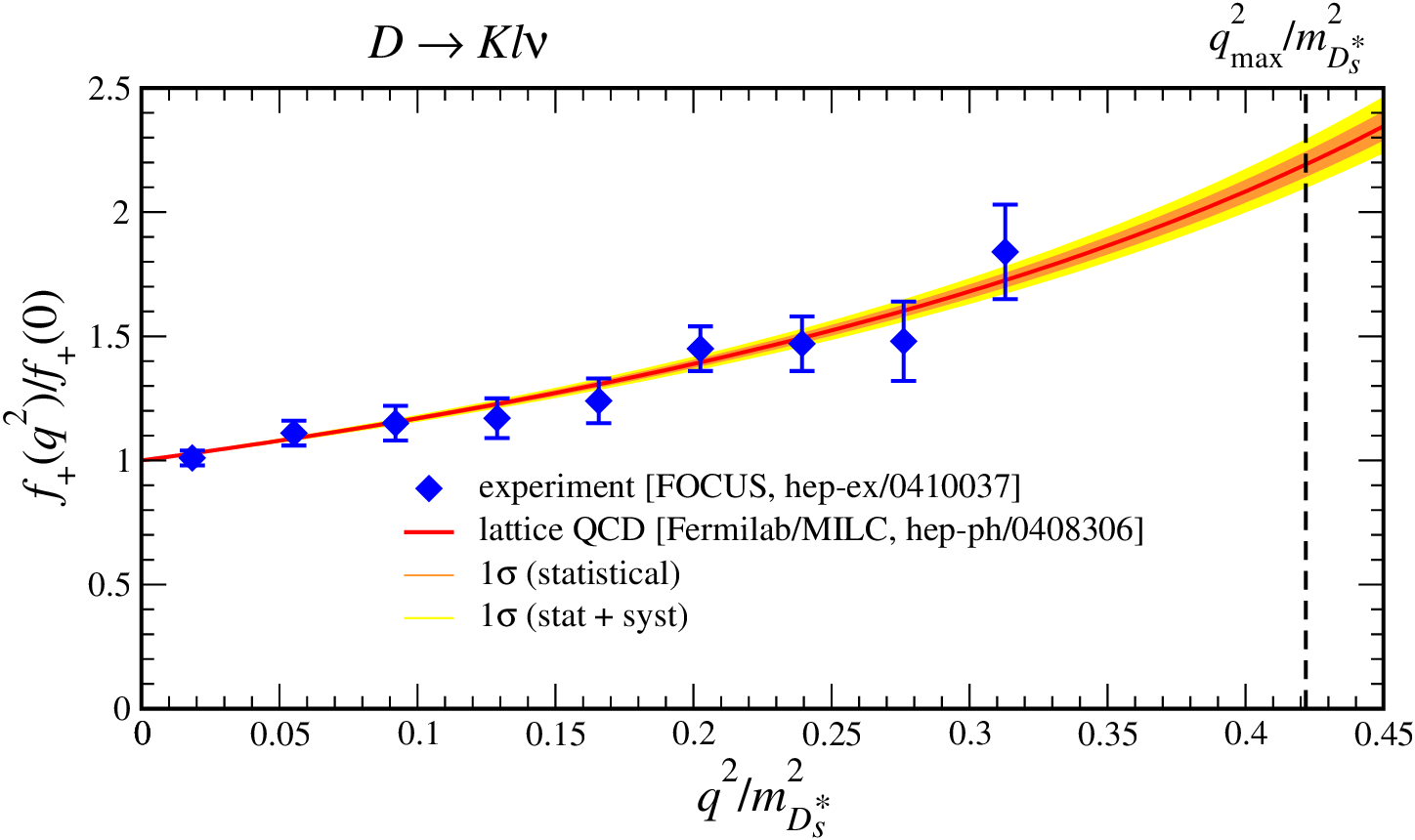} \hfill
	(b)\hspace*{-1.0em}\includegraphics[width=0.45\textwidth]{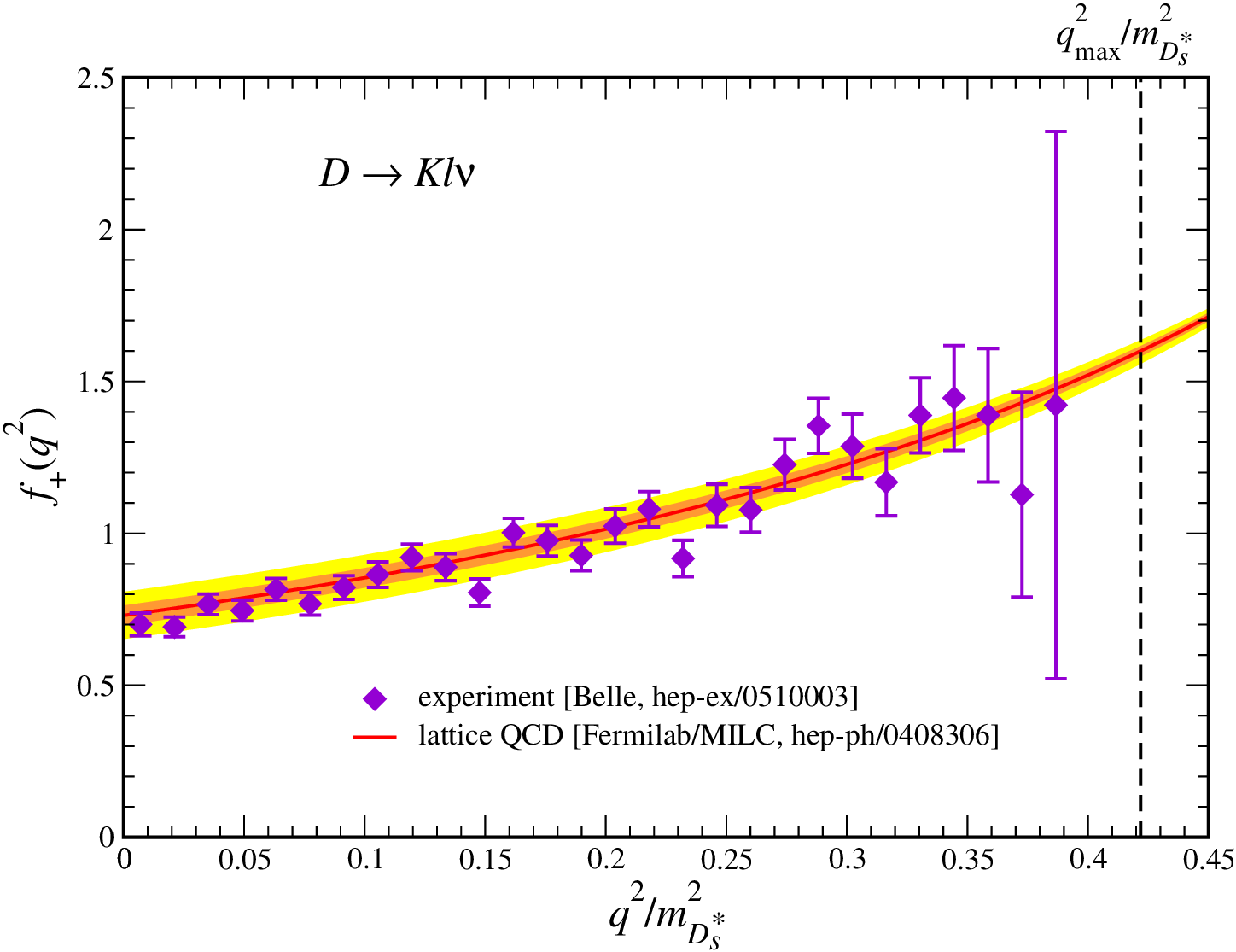}
	\caption[fig:focus]{Form factor for $D\to Kl\nu$ vs.~$q^2/m^2_{D^*_s}$:
	(a) shape $f_+(q^2)/f_+(0)$	compared with FOCUS~\cite{Link:2004dh};
	(b) shape and normalization $f_+(q^2)$ compared with 
	Belle~\cite{Widhalm:2006wz}.}
	\label{fig:focus}
\end{figure}

In Fig.~\ref{fig:focus}(a) we plot our result for $f_+(q^2)/f_+(0)$
vs.~$q^2/m^2_{D^*_s}$.
The errors from $f_+(0)$ must be propagated to non-zero~$q^2$,
so for $f_+(q^2)/f_+(0)$ the errors grow with~$q^2$.
Figure~\ref{fig:focus} shows 1-$\sigma$ bands of statistical (orange)
and all uncertainties (yellow) added in quadrature~\cite{Mackenzie:2005di}.
As one can see, the $q^2$ dependence of lattice QCD (curve and error
band) and data from the FOCUS experiment (points) agree excellently, although
the uncertainties are still several per cent.
The FOCUS results appeared two months after the lattice calculation.
More recently, the Belle Collaboration at the $e^+e^-$ collider KEK-$B$
measured the shape \emph{and} normalization of the form factor in a
single experiment~\cite{Widhalm:2006wz}.
In Fig.~\ref{fig:focus}(b) we compare our result for $f_+(q^2)$ with 
Belle.
The color code for the lattice QCD error bands is as before, and now 
depict $q^2$ dependence of the lattice-QCD errors in a realistic~way.

%% file: 3.tex
We also considered the leptonic decay of charmed mesons, $D^+\to l\nu$
and~$D_s\to l\nu$.
Here the quark and antiquark in the meson merge into a virtual $W$, 
which disintegrates into~$l\nu$.
The QCD influence is a single number (for each meson),
called decay constants and denoted~$f_{D^+}$ or~$f_{D_s}$.
At Lattice 2004 \cite{Simone:2004fr}, we presented preliminary results
for $f_{D^+}$ and $f_{D_s}$, based on one lattice spacing,
$a\approx0.125$~fm.
Our extended the running to two other lattice spacings.
Details are given in the ensuing publication~\cite{Aubin:2005ar}.
We find
\begin{eqnarray}
	f_{D^+} &\!\! =\!\! & 201 \pm 3 \pm 17~\textrm{MeV},
	\label{eq:fD+:lat} \\
	f_{D_s} &\!\! =\!\! & 249 \pm 3 \pm 16~\textrm{MeV},
	\label{eq:fDs:lat}
\end{eqnarray}
where the first error is from finite Monte Carlo statistics, the second
is a sum in quadrature of several systematics.
A~conservative (but not na\"ive) estimate of heavy-quark discretizations
effects is the second largest (largest) systematic on $f_{D^+}$
($f_{D_s}$).

\begin{figure}[b]
	\centering
	(a) \includegraphics[height=1.75in]{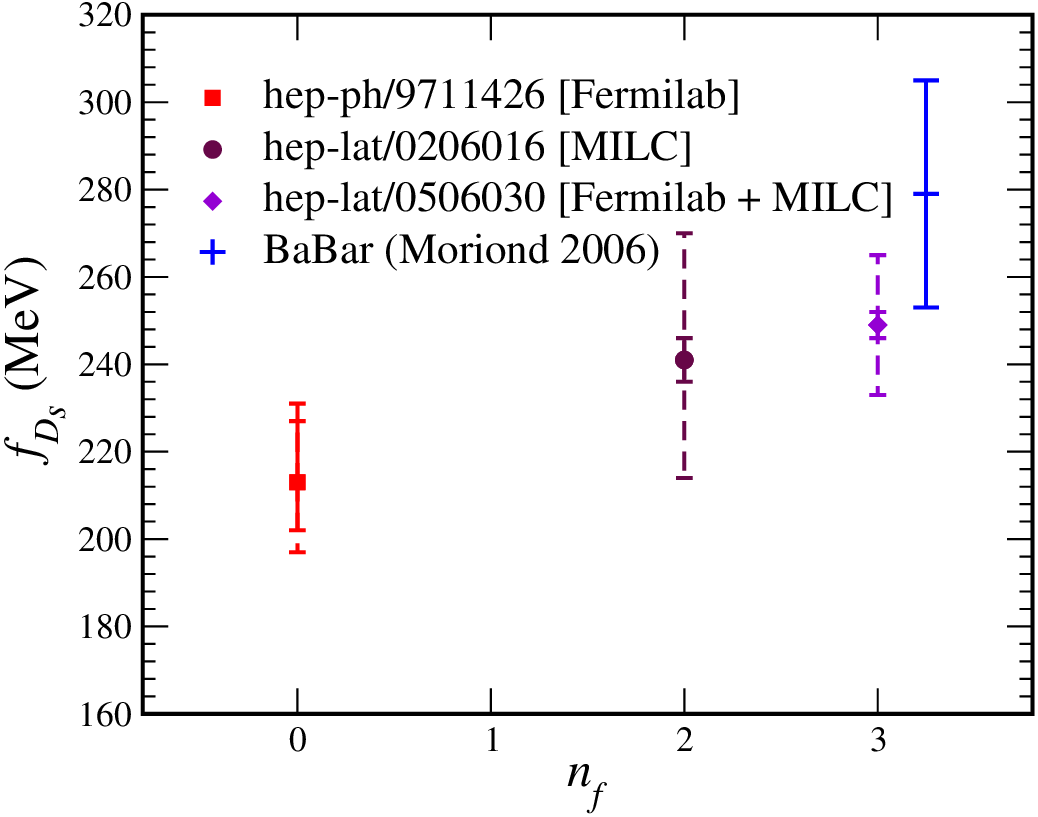}\hfill
	(b) \includegraphics[height=1.75in]{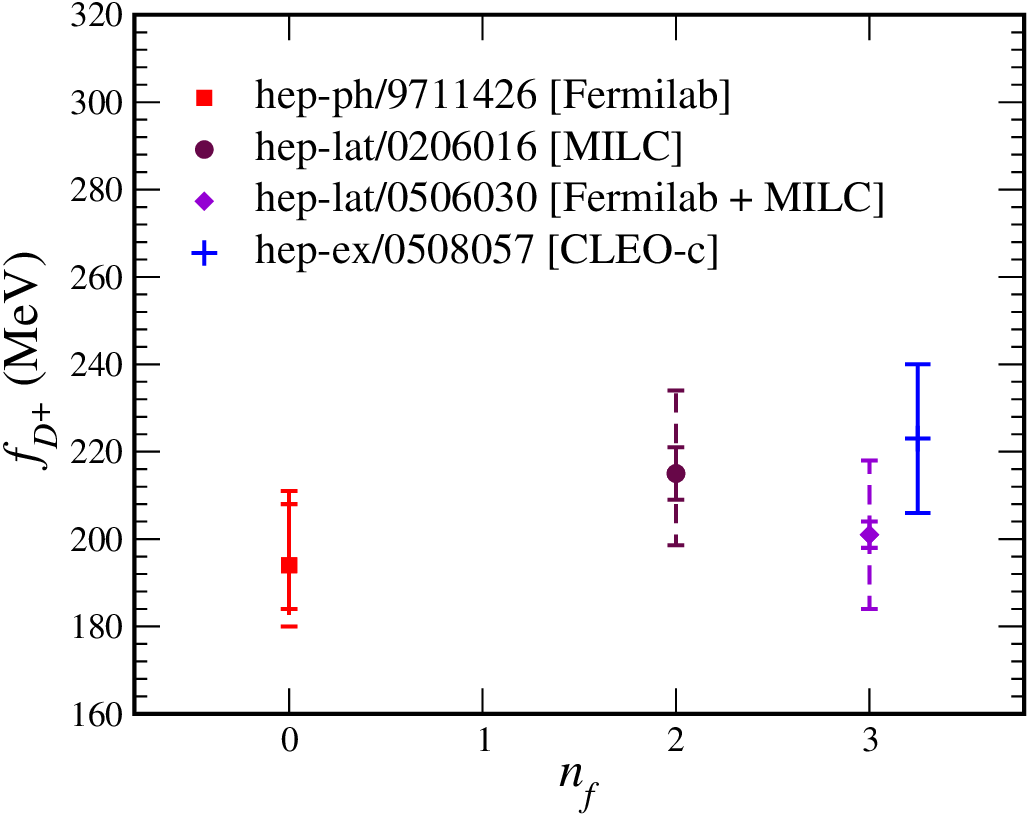}
	\caption[fig:mBc]{Dependence of (a) $f_{D_s}$ and (b) $f_{D^+}$
	on the number $n_f$ of sea flavors.
	Quenched ($n_f=0$)~\cite{El-Khadra:1997hq};
	$n_f=2$~\cite{Bernard:2002pc};
	$n_f=3$~\cite{Aubin:2005ar}.
	Solid (dashed) error bars are statistical (statistical+systematic).}
	\label{fig:fD}
\end{figure}

Figure~\ref{fig:fD} shows the $n_f$ dependence of the decay constants.
Quenched ($n_f=0$) results vary widely, but we show
one~\cite{El-Khadra:1997hq} carried out with similar choices for other 
aspects of the calculations.
One sees a trend of $f_{D_s}$ to increase with $n_f$.
A~similar comparison of $f_{D^+}$, in Fig.~\ref{fig:fD}(b), is less
instructive, but shown for completeness.

The CLEO-$c$ Collaboration~\cite{Artuso:2005lp} and 
the BaBar Collaboration~\cite{Berryhill:2006mo} have measured
\begin{eqnarray}
	f_{D^+} &\!\! =\!\! & 223 \pm 17 \pm 3~\textrm{MeV}
		\quad \textrm{CLEO-}c~\cite{Artuso:2005lp}, \label{eq:fD+:expt} \\
	f_{D_s} &\!\! =\!\! & 279 \pm 17 \pm 20~\textrm{MeV}
		\quad \textrm{BaBar}~\cite{Berryhill:2006mo}, \label{eq:fDs:expt}
\end{eqnarray}
respectivelty at the CESR and PEP-II $e^+e^-$ colliders.
At the 1-$\sigma$ level, the agreement with lattice QCD is fine.
Even more compelling is the ratio
$R_{d/s} = m^{1/2}_{D^+}f_{D^+}/m^{1/2}_{D_s}f_{D_s}$:
\begin{eqnarray}
	R_{d/s} &\!\! =\!\! & 0.786 \pm 0.042 \quad \textrm{lattice QCD}, \\
			&\!\! =\!\! & 0.779 \pm 0.093 \quad \textrm{CLEO/BaBar},
\end{eqnarray}
in which several uncertainties from lattice QCD cancel.
Experimental and lattice-QCD uncertainties are reducible, so the test 
will sharpen over the coming few years.

%% file: 4.tex
The pseudoscalar $B_c^+$ meson is the lowest-lying bound state of a
charmed quark and a $b$ quark.
The CDF Collaboration~\cite{Abe:1998wi} first observed it during Run~I
of the Tevatron $\bar{p}p$ collider in the semileptonic decay 
$B_c^+\to J/\psi l^+\nu$.
But it was clear that Tevatron Run~II detectors would be able to
reconstruct hadronic modes, such as $B_c^+\to J/\psi\pi^+$, which give
much much better precision on~$m_{B_c}$ \cite{Anikeev:2001rk}.
At Lattice 2004 we presented results in nearly final
form~\cite{Allison:2004hy}, and posted the final results on the arXiv
in mid-November~\cite{Allison:2004be}:
\begin{equation}
	m_{B_c} = 6304 \pm 12^{+18}_{-~0}~\textrm{MeV},
	\label{eq:mBc:lat}
\end{equation}
where the last error is a rough estimate of residual heavy-quark
discretization effects.
Soon afterwards, CDF announced their precise mass
measurement~\cite{Acosta:2005us}:
\begin{equation}
	m_{B_c} = 6287 \pm 5~\textrm{MeV},
	\label{eq:mBc:expt}
\end{equation}
which agrees with Eq.~(\ref{eq:mBc:lat}) at slightly more than
1-$\sigma$.

\begin{figure}[b]
	\centering
	(a) \includegraphics[height=0.2\textwidth]{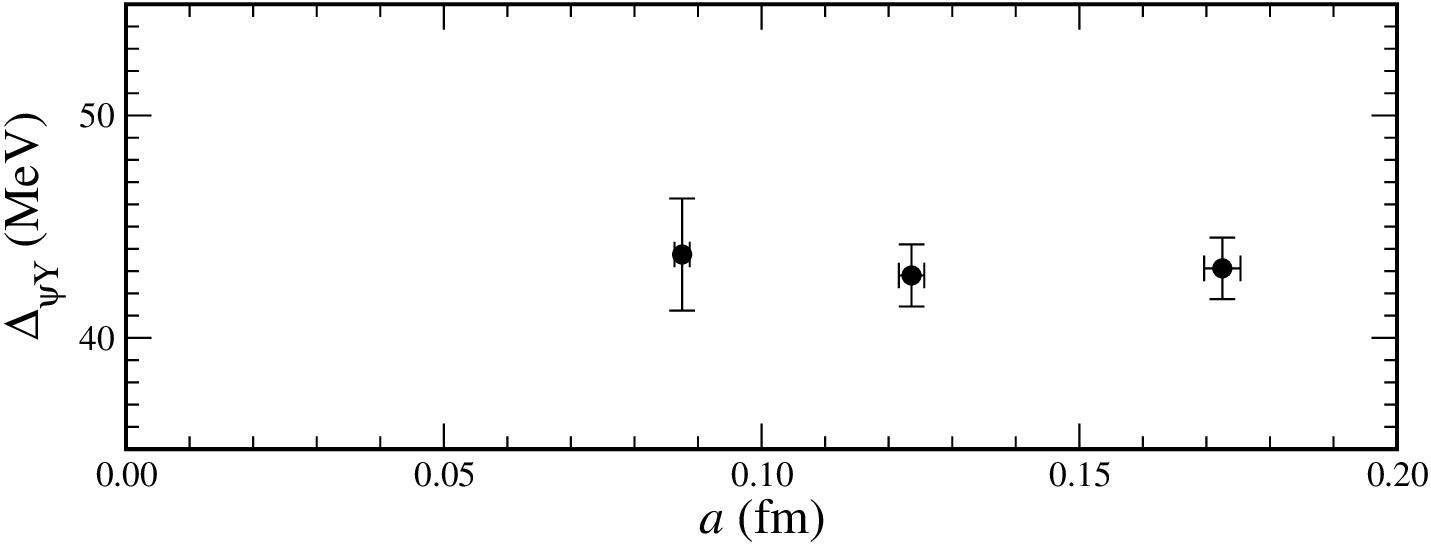}\hfill
	(b) \includegraphics[height=0.2\textwidth]{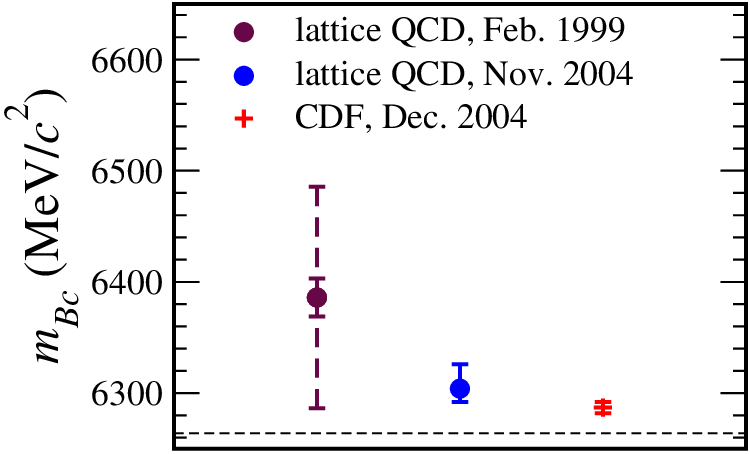}
	\caption[fig:mBc]{(a) Dependence of the
	splitting~$\Delta_{\psi\Upsilon}$ on the lattice spacing~$a$.
	(b) Comparison of the quenched~\cite{Shanahan:1999mv},
	$n_f=2+1$~\cite{Allison:2004be} and
	experimental~\cite{Acosta:2005us} values of $m_{B_c}$;
	the dashed line denotes the baseline
	$(\bar{m}_\psi+m_\Upsilon)/2$.}
	\label{fig:mBc}
\end{figure}

Two comments are in order.
First, the agreement at the gross level of the calculation with
experiment shows that discretization effects are well under control
with the heavy-quark methods of choice.
These are lattice NRQCD~\cite{Lepage:1987gg} and the Fermilab
method~\cite{El-Khadra:1996mp}, which are based on effective field
theories for heavy quarks~\cite{Lepage:1992tx,Kronfeld:2000ck}.
Indeed, as seen in Fig.~\ref{fig:mBc}(a), almost no lattice spacing
dependence is seen in the splitting
$\Delta_{\psi\Upsilon}=m_{B_c}-(\bar{m}_\psi+m_\Upsilon)/2$ that is at
the crux of the calculation~\cite{Shanahan:1999mv}.
Moreover, it is striking how little the splitting~$\Delta_{\psi\Upsilon}$
changes when sea quarks are included.
Figure~\ref{fig:mBc}(b) compares Eq.~(\ref{eq:mBc:lat}) with an old
quenched calculation~\cite{Shanahan:1999mv} (and the
measurement~\cite{Acosta:2005us}).
The solid error bar shows the non-quenching errors,
and the dashed includes the estimate of the quenching error.
The inclusion of sea quarks has reduced the splitting by a factor of
three or four, bringing an essentially discrepant result into agreement.

%% file: 5.tex
In the past year, several lattice-QCD calculations have been confirmed by
experiment.
FOCUS \cite{Link:2004dh} and Belle~\cite{Widhalm:2006wz} confirmed the
$q^2$-dependence of the $D\to Kl\nu$ form factor~\cite{Aubin:2004ej};
CLEO-$c$~\cite{Artuso:2005lp} and BaBar~\cite{Berryhill:2006mo}
respectively confirmed the $D^+$ and $D_s$ decay
constants~\cite{Aubin:2005ar};
and CDF~\cite{Acosta:2005us} confirmed the mass of the $B_c$
meson~\cite{Allison:2004be}.
To obtain these results it is essential to have heavy-quark
discretization effects under control, as one expects from theoretical
foundations~\cite{Lepage:1987gg,El-Khadra:1996mp,Lepage:1992tx,Kronfeld:2000ck}.
Furthermore, the comparison of quenched QCD, QCD with 2+1 staggered
flavors, and experiment shows that sea quarks are needed to obtain
agreement, and that staggered quarks (in these cases) capture the needed
effect.
The results are promising for the search for new $b$ quark 
interactions, because a straighforward change to the $D$ form factors 
and decay constants yield the corresponding results for $B$ mesons.
These are a key element to enable the experimental search for new
phenomena in quark-flavor physics~\cite{Shipsey:2004wz}.